\documentclass[conference]{IEEEtran}
\usepackage{cite}
\usepackage{amsmath,amssymb,amsfonts}
\usepackage{algorithmic}
\usepackage{graphicx}
\usepackage{textcomp}
\usepackage{xcolor}
\PassOptionsToPackage{hyphens}{url}\usepackage{hyperref}
\graphicspath{ {./images/} }
\def\BibTeX{{\rm B\kern-.05em{\sc i\kern-.025em b}\kern-.08em
    T\kern-.1667em\lower.7ex\hbox{E}\kern-.125emX}}
    
\begin{document}

\title{Study of Firecracker MicroVM}

\author{\IEEEauthorblockN{Madhur Jain}
\IEEEauthorblockA{\textit{Khoury College of Computer Sciences} \\
\textit{Northeastern University}\\
Boston, MA\\
jain.madh@northeastern.edu}
}

\maketitle


\begin{abstract}
Firecracker is a virtualization technology that makes use of Kernel Virtual Machine (KVM). Firecracker belongs to a new virtualization class named the micro-virtual machines (MicroVMs). Using Firecracker, we can launch lightweight MicroVMs in non-virtualized environments in a fraction of a second, at the same time offering the security and workload isolation provided by traditional VMs and also the resource efficiency that comes along with containers \cite{b1}. Firecracker aims to provide a slimmed-down MicroVM, comprised of approximately 50K lines of code in Rust and with a reduced attack surface for guest VMs. This report will examine
the internals of Firecracker and understand why Firecracker is the next big thing going forward in virtualization and cloud computing.
\end{abstract}


\section{Introduction}
Firecracker is a new open source Virtual Machine Monitor (VMM) developed by AWS for serverless workloads. It 
is a virtualization technology which makes use of KVM, meaning it can only be run on KVM-supported and enabled hosts and a corresponding Linux Kernel v4.14+. With the recommended Linux kernel guest configuration, Firecracker claims to offer a memory overhead of less than 5MB per container, boots to application code within 125ms, and allows for the creation of up to 150 MicroVMs per second. The number of Firecracker MicroVMs running simultaneously on a single host is only limited by the availability of hardware resources. \cite{b1}

Firecracker provides security and resource isolation by running the Firecracker userpsace process inside a jailer process. The jailer sets up system resources that require elevated permissions (e.g., cgroup, chroot), drops privileges, and then exec()s into the Firecracker binary, which then runs as an unprivileged process. Past this point, Firecracker can only access resources to which a privileged third-party grants access (e.g., by copying a file into the chroot, or passing a file descriptor). \cite{b13}

Seccomp filters limit the system calls that the Firecracker process can use. There are 3 possible levels of seccomp filtering, configurable by passing a command line argument to the jailer: 0 (disabled), 1 (whitelists a set of trusted system calls by their identifiers) and 2 (whitelists a set of trusted system calls with trusted parameter values), the latter being the most restrictive and the recommended one. The filters are loaded in the Firecracker process, immediately before the execution of the untrusted guest code starts. \cite{b13}

Firecracker was developed to handle serverless workloads and has been running in production for AWS Lambda and Fargate since 2018. Serverless workloads require isolation and security, and at the same time, have container benefits of faster boot time. A lot of research has been done with respect to the cold-start and warm-start of the VM instances for serverless workloads. \cite{b17} The idea behind developing Firecracker was to make use of the KVM module loaded in the Linux kernel and get rid of the legacy devices that other virtualization technologies like Xen and VMWare offer. This way, Firecracker can create a VMM with a smaller memory footprint and also provide improved performance.

Section 2 explains the high-level architecture of the Firecracker Micro VM, Section 3 dives deep into the boot sequence of the Firecracker, Section 4 describes the device model emulation provided, and Section 5 shares some light on the conclusions garnered through this study.

\section{Architecture of Firecracker}

KVM is an enabler of hardware extensions provided by vendors such as Intel and AMD with their virtualization extensions such as SVM and VMX. These extensions allow KVM to directly execute the guest code on the host CPU. There are three sets of ioctls that make up the KVM API and are issued to control the various aspects of the virtual machine. The three classes that the iocltls belongs to are \cite{b7} -

\begin{itemize}
    \item \textbf{System IOCTLs}: These query and set global attributes, which affect the whole KVM subsystem. In addition, a system ioctl is used to create virtual machines.
    \item \textbf{VM IOCTLs}: These query and set attributes that affect an entire virtual machine—for example, memory layout. In addition, a VM ioctl is used to create virtual CPUs (vCPUs). VM ioctls are run from the same userspace process (address space) that was used to create the VM.
    \item \textbf{vCPU IOCTLs}: These query and set attributes that control the operation of a single virtual CPU. They run vCPU ioctls from the same thread that was used to create the vCPU.
\end{itemize}

\begin{figure}
    \centering
    \includegraphics[width=8cm, height=4cm]{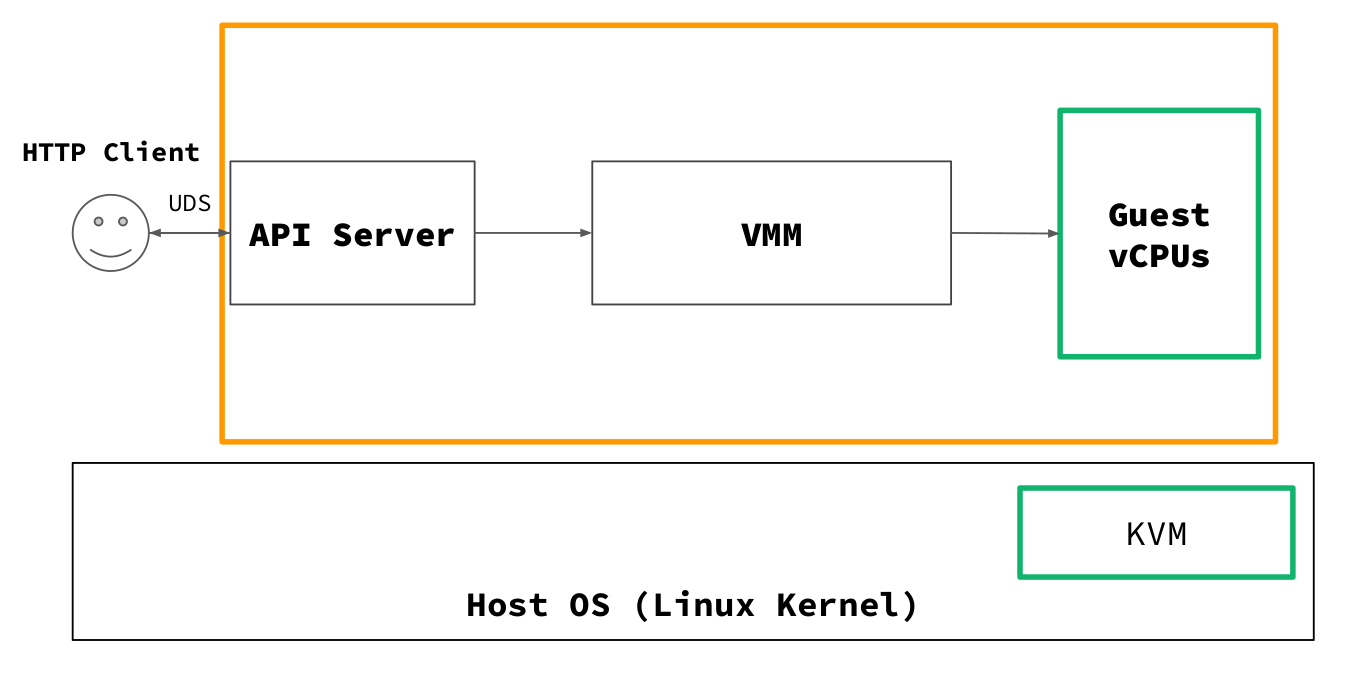}
    \caption{Firecracker Design - Threads}
    \label{fig:threads}
\end{figure}{}

QEMU is an open source machine emulator and a virtualization technique to run KVM-enabled Virtual Machines. Similar to QEMU, Firecracker uses a multithreaded event-driven architecture. Each Firecracker process runs one and only one MicroVM. The process consists of three main threads: \textbf{API, VMM and vCPU}. The API server thread is an HTTP server that is created to accept requests and performs actions on those requests. The API server thread structure comprises of a socket connection to listen to requests on the port, epoll\_fd to listen to events on the socket port, and a hashmap of connections. The hashmap consists of token and connection pairs corresponding to the file descriptor of the stream—in this case, the socket.


Traditionally, QEMU has made use of \textit{select} or \textit{poll} system calls to maintain an event loop of file descriptors on which to listen for new events. \cite{b16} \textit{select} or \textit{poll} system calls requires a list of all open file descriptors maintained in the VMM structure and then it
would go through each of the file descriptors to determine which FDs have new events. This would take up O(N) time where N is the number of file descriptors to listen for new events. \cite{b14}


Firecracker takes the epoll approach where the host kernel maintains a list of file descriptors for VM process and notifies the VMM whenever there is a new event that occurs in any of the file descriptors. This is called as a Edge Triggered mechanism ("pull"), whereas the select/poll was a Level Triggered mechanism ("push"). The epoll\_fd structure created by Firecracker has the `close\_on\_exec` flag set, which means if a process forks the Firecracker process, the file descriptors would not be shared.


The API server exposes a number of requests as REST APIs which can be used to configure the MicroVM. Once the MicroVM has been started, i.e., once it receives the "InstanceStart" command, the API server thread will just block on the epoll file descriptor until new events come in. Firecracker creates a channel (see Rust channels) to enable communication between the API Server thread and the VMM thread. Rust channels are similar to Unix pipes for comparison.

The VMM server thread manages the entire MicroVM. Once the VMM server thread is created, it runs an event loop which takes the parsed request one by one from the API server thread and dispatches it to the appropriate handlers. The handlers are defined according to the dispatch table set by the event loop. For now, Firecracker supports the following handlers - Exit, Stdin, DeviceHandler, VMMActionRequest, WriteMetrics. The dispatch table is managed by the epoll fd. The dispatch table maintains a map of file descriptors that are to be monitored, and the kind of events to be monitored for. When the vCPU thread creation request is received by the VMM thread, the VMM spawns the required number of vCPU by using the KVM vCPU ioctls. A vCPU thread is created for each virtual CPU. These vCPU threads are nothing but POSIX threads created by KVM. To run guest code, the vCPUs execute an ioctl with KVM\_RUN as its argument.

Software in the root mode is the hypervisor. Hypervisor or the VMM forms a new plane that runs in root mode while the VMs run in non-root mode. KVM uses virtualization extensions to provide these different modes on the host CPUs. In the case of Intel CPUs, VT-x is the CPU virtualization and VT-d is the IO virtualization. For vCPUs, VT-x provides two modes of guest code execution: root and non-root. Whenever a VM attempts to execute an instruction that is prohibited by the non-root model, vCPU immediately switches to a root mode in a trap-like way.
This is called a VMEXIT. Hypervisor deals with the reason for VMEXIT and then executes VMRESUME to re-enter non-root mode for that VM instance. This interaction between root and non-root is the essence of hardware virtualization. 



\section{Firecracker Boot Sequence}
Traditional PCs boot Linux with a BIOS and a bootloader. The primary responsibilities of the BIOS includes booting the CPU in real mode and performing a Power on Self Test (POST) setup before loading the bootloader. BIOS determines the candidates for boot devices, and once a bootable device is found, the bootloader is loaded into RAM and executed. Different systems have different stages of the bootloader to be executed. LILO, for example, has a two stage bootloader while GRUB contains a 3 stage bootloader. Multiple stages of a bootloader are used because of the system limitations of some of the older devices that were used to boot Linux.

Linux kernels actually do not require a BIOS and a bootloader. Instead, Firecracker uses what is known as \textbf{Linux Boot Protocol}. \cite{b15} There are multiple versions of the Linux Boot Protocol standard that exist. Firecracker follows the 64-bit Linux Boot Protocol Standard. Thus, Firecracker can directly load the Linux kernel and mount the corresponding root file system.


The Linux kernel is an uncompressed \textit{bzImage} (big compressed image, usually larger than 512KB). The \textit{bzImage} format consists of a real mode kernel code and a protected mode kernel code. Instead of booting into the entry point defined by the real mode, Firecracker directly boots into the 64-bit entry point located at 0x200 in the protected mode kernel code. Firecracker loads the uncompressed Linux kernel as well as the init process, thereby saving approximately 20 to 30ms of the time taken to decompress the kernel.



\begin{figure*}
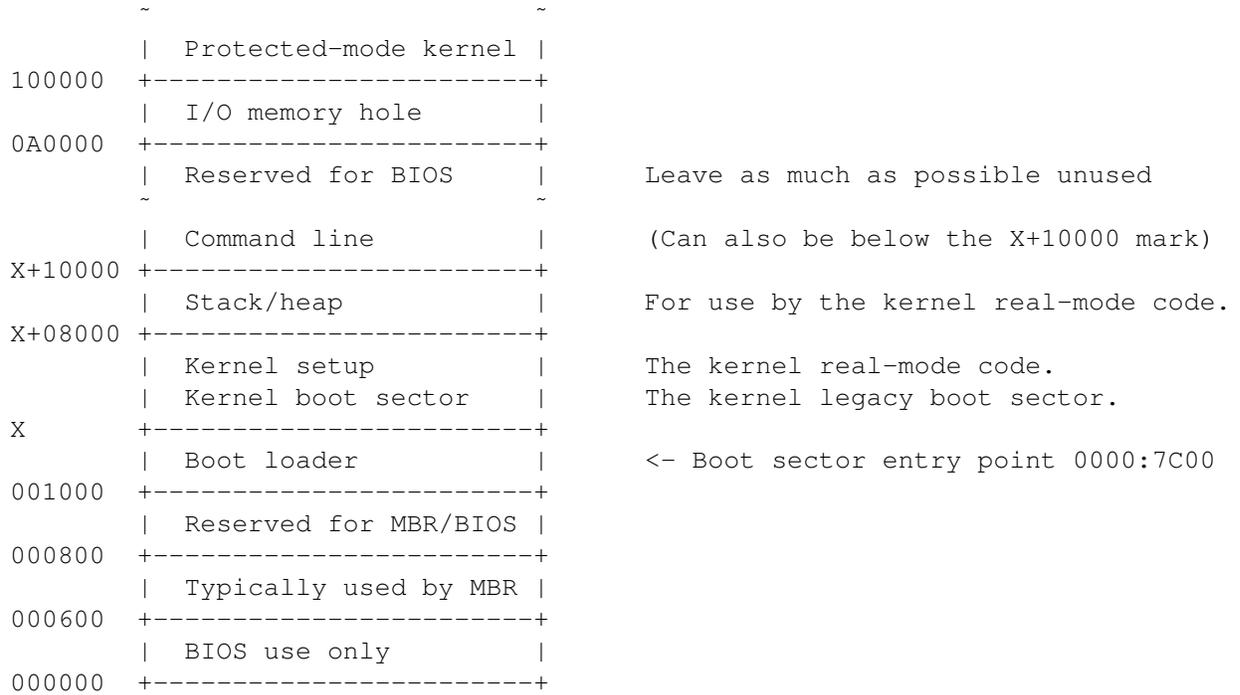

\centering
\begin{verbatim}
              ~                        ~
              |  Protected-mode kernel |
      100000  +------------------------+
              |  I/O memory hole       |
      0A0000  +------------------------+
              |  Reserved for BIOS     |      Leave as much as possible unused
              ~                        ~
              |  Command line          |      (Can also be below the X+10000 mark)
      X+10000 +------------------------+
              |  Stack/heap            |      For use by the kernel real-mode code.
      X+08000 +------------------------+
              |  Kernel setup          |      The kernel real-mode code.
              |  Kernel boot sector    |      The kernel legacy boot sector.
      X       +------------------------+
              |  Boot loader           |      <- Boot sector entry point 0000:7C00
      001000  +------------------------+
              |  Reserved for MBR/BIOS |
      000800  +------------------------+
              |  Typically used by MBR |
      000600  +------------------------+
              |  BIOS use only         |
      000000  +------------------------+
\end{verbatim}
\caption{Linux BzImage Memory Layout}
\label{fig:bzimage}
\end{figure*}


Linux kernel also contains another component, namely the initramfs \cite{b5}. There are four primary reasons to have an initramfs in the LFS environment: loading the rootfs from a network, loading it from an LVM logical volume, having an encrypted rootfs where a password is required, or for the convenience of specifying the rootfs as a LABEL or UUID. Anything else usually means that the kernel was not configured properly.


Since Firecracker doesn't need any of the above stated reasons for loading the initramfs before mounting the root file system, it is recommended to avoid loading the initramfs at boot time, thereby further reducing the overall boot time and the memory footprint for the kernel. So, if no initramfs is configured externally, then at boot time, Firecracker replaces the initramfs with a default empty, 134 byte initramfs.

\section{Firecracker Device Model}
Until this section, we have talked about the similar architectures and execution flow for Firecracker and QEMU. So, what is different between QEMU and Firecracker? One of the main differences is with the device emulations. There are only 5 Device emulations available in Firecracker: network, block devices, sockets, serial console and minimal keyboard controller, as shown in Figure 3. Firecracker does not provide support for device emulations like USB, GPU and 9P filesystem in order to provide increased security compared to other virtualization technologies like QEMU. On the other hand, QEMU has most device model emulations available in the VMM. The careful reader will notice that Firecracker does not the use the vhost implementation in the host kernel that provides more efficient IO performance without doing VMEXITS.

An open specification for emulating device models in virtualization has been developed, named Virtio. Virtio is defined as a straightforward, efficient and a standard mechanism to allow guest OS to talk to the virtual device driver in a similar way the host OS would call the actual hardware device driver. It takes advantage of the fact that the guest can share memory with the host for IO.


The general flow for the virtio specification \cite{b8} includes a front-end driver representing the virtual device in the guest, and the corresponding device being exposed by the hypervisor or the VMM. A transport layer enables communication between the host and the guest. For the transport layers, Virtio employs a ring-buffer virtqueue structure. A virtuqueue is a queue of guest-allocated buffers that the host interacts with either by reading or writing to them. Each device can have zero or more virtqueues. A back-end driver present in the host kernel completes the communication flow, to which the virtqueue is connected.

Firecracker device model architecture using Virtio is shown in Figure 3. The following list provides a description of the devices available within Firecracker: 
\begin{itemize}
    \item \textbf{virtio-net}: implementation for the network driver (tun/tap devices)
    \item \textbf{virtio-blk}: implementation for the block devices
    \item \textbf{virtio-vsock}: implementation for VM sockets providing N:1 serial communications
    \item \textbf{serial console}: implementation for the legacy console devices for serial communication - terminal
    \item \textbf{keyboard controller}: implementation for the keyboard device, though only one function is implemented - Ctrl+Alt+Del to reboot/shutdown the system.
\end{itemize}

\begin{figure}
    \centering
    \includegraphics[width=8cm, height=4cm]{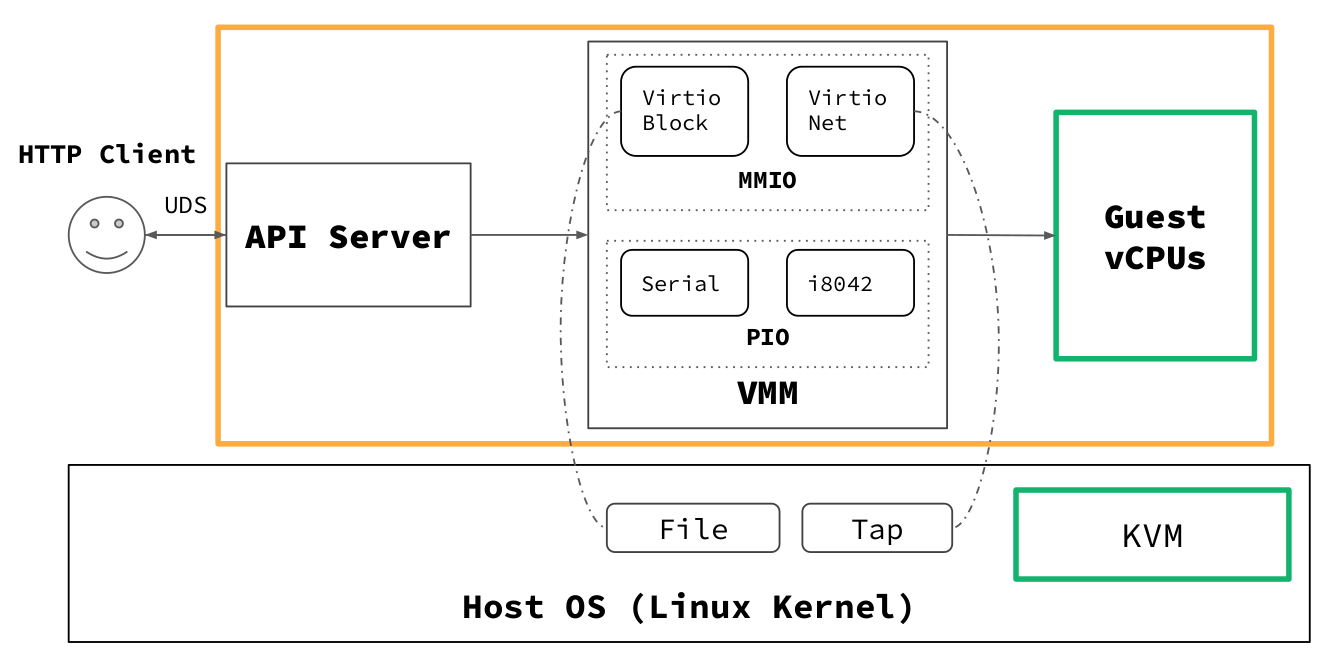}
    \caption{Firecracker Device Model}
    \label{fig:device_model}
\end{figure}{}

\section{Scope for Improvements}
Even with all the excellent features providing near-native performance of the guest code using KVM, as well as faster boot times and lower memory footprint of the VMs due to fewer support for available device emulations, there still exist some areas for improvements that can make Firecracker more suitable for general use cases and not just for serverless workloads.
\begin{itemize}
    \item \textbf{Support for virtio-fs}: virto-fs is the interface to provide efficient sharing between the host and the guest filesystem avoiding context switches (VMEXITS) thereby providing more performance. virtio-fs is an upgrade on the existing virtio-9p interface for the same purpose. Though more research is required for security purposes before including it as part of Firecracker.
    \item \textbf{Increased IO Performance}: The results of the tests performed between the Firecracker, QEMU and Cloud Hypervisor show limitations in Firecracker's virtio implementation and serial execution. \cite{b1} \cite{b2}
    \item \textbf{Larger number of device emulations}: Currently, Firecracker can emulate only 10 devices, since each device gets its own IRQ. \cite{b10}
    \item \textbf{Support for attaching devices at runtime}: Firecracker only allows specifying the devices at booting time. Devices can only be attached when the MicroVM is shut down.
    \item \textbf{Hotplugging Support}: For any workload, it is beneficial to allow guest memory/CPU hotplugging within a VM at runtime in order to avoid interference to the workload. Firecracker oversubscribes the allocated memory required for the guest, but there is no way to expand the allocated memory for the guest.
    \item \textbf{Memory Ballooning Support}: At present, Firecracker does not have any support for reclaiming unused memory from the guest, since no communication is present between the host and the guest. This, along with the hotplugging feature would make it very easy to dynamically add/remove memory/CPU at runtime thereby providing elasticity to the MicroVM. \cite{b9}
\end{itemize}{}

\section{Conclusion/Thoughts}
This paper reviews the implementation of a minimalist and modular VMM in the form of Firecracker MicroVM. It also identifies the process of how Firecracker provides resource isolation and security through the use of seccomp filters and jailer process and provides faster boot times and lower memory footprint due to KVM and minimal device model emulation.


One other thing to note is that Firecracker embodies
the modular design in the development of the hypervisor. The modular design approach has also led to the development of community-driven high-quality \textbf{rust-vmm} crates which provide us with the core modules required for the implementation of a hypervisor \cite{b11}. rust-vmm \cite{b12} is a community approach initiated by AWS. Amazon along with Intel, Redhat and Google, are trying to provide a platform to build a hypervisor from scratch by only consuming the modules required from the rust-vmm crates. This approach also enables the development of a plug-n-play architecture in hypervisors, which we haven't seen so far.

\bibliography{references}
\end{document}